\documentclass[manuscript]{emulateapj}

\newcommand{\msun}{M$_\odot$} 
 
\newcommand{\kms}{\,\hbox{km}\,\hbox{s}^{-1}}

\def\nar{NewAR} 
\def\simgt{\lower.5ex\hbox{\gtsima}} 
\def\simlt{\lower.5ex\hbox{\ltsima}} 
 
\shorttitle{Storm in a ``Teacup''} 
\shortauthors{Harrison et al.} 
 
\begin{document} 
 
\title{Storm in a ``Teacup'': a radio-quiet quasar with $\approx$10\,kpc radio-emitting bubbles and extreme gas kinematics} 
 
\author{
C.\ M.\ Harrison,\altaffilmark{1,$\star$}
A.\ P.\ Thomson,\altaffilmark{1}
D.\ M.\ Alexander,\altaffilmark{1}
F.\ E.\ Bauer,\altaffilmark{2,3,4}
A.\ C.\ Edge,\altaffilmark{1}
M.\ T. Hogan,\altaffilmark{1}
J.\ R.\ Mullaney\altaffilmark{5}
and A.\ M.\ Swinbank\altaffilmark{1}
}
\altaffiltext{$\star$}{Email: c.m.harrison@mail.com}
\altaffiltext{1}{Department of Physics, Durham University, South Road,
  Durham, DH1 3LE, U.K.}
\altaffiltext{2}{Instituto de Astrof\'{i}sica, Facultad de F\'{i}sica,
  Pontifica Universidad Cat\'{o}lica de Chile, 306, Santiago 22,
  Chile}
\altaffiltext{3}{Millennium Institute of Astrophysics, Vicu\~{n}a
  Mackenna 4860, 7820436 Macul, Santiago, Chile}
\altaffiltext{4}{Space Science Institute, 4750 Walnut Street, Suite 205, Boulder, CO 80301, USA}
\altaffiltext{5}{Department of Physics and Astronomy, University of
  Sheffield, Sheffield, S7 3RH, U.K.}


\begin{abstract} 
We present multi-frequency (1--8\,GHz) VLA data, combined with VIMOS IFU data and {\em HST} imaging, of
a $z=0.085$ radio-quiet type~2 quasar (with $L_{{\rm
    1.4\,GHz}}\approx5\times10^{23}$\,W\,Hz$^{-1}$ and $L_{{\rm
    AGN}}\approx2\times10^{45}$\,erg\,s$^{-1}$).  Due to
the morphology of its emission-line region, the target
(J1430+1339) has been referred to as the Teacup AGN in the literature. We identify
``bubbles'' of radio emission that are extended $\approx$10--12\,kpc to both the east and
west of the nucleus. The edge of the brighter eastern bubble is co-spatial
with an arc of luminous ionized gas. We also show that the Teacup AGN
hosts a compact radio structure, located $\approx$0.8\,kpc from the core position, at the base of the
eastern bubble. This radio structure is co-spatial with an ionized outflow with an
observed velocity of $v=-740$\,km\,s$^{-1}$. This is likely to correspond to a jet, or possibly a quasar wind, interacting with the
interstellar medium at this position. The large-scale radio bubbles
appear to be inflated by the central AGN, which indicates that the AGN can also interact with
the gas on $\gtrsim10$\,kpc scales. Our study highlights that even
when a quasar is formally ``radio-quiet" the radio emission can be extremely
effective for observing the effects of AGN feedback. 
\end{abstract} 
 
 

\section{Introduction} 
There is now general agreement that growing black holes, i.e., active galactic
nuclei (AGN) activity, can have a profound impact upon the gas in their host galaxies and
beyond. This impact, often termed ``AGN feedback'', is a
requirement of most models of galaxy formation in order to reproduce the
properties of local massive galaxies, the intergalactic medium and
intracluster medium (e.g., BH-mass scaling relationships; the galaxy
luminosity function; the X-ray temperature-luminosity relationship;
e.g., \citealt{Silk98}; \citealt{Benson03}; \citealt{Churazov05};
\citealt{Hopkins06}; \citealt{Bower06}; \citealt{McCarthy10};
\citealt{Gaspari11}; \citealt{Schaye14}; \citealt{Vogelsberger14}). Observationally, there is
convincing evidence that some sort of interaction occurs, although
establishing the types of galaxies in which it is important and exactly how the energy couples to the gas is a matter of ongoing
research (see reviews in \citealt{Alexander12}; \citealt{Fabian12};
\citealt{McNamara12}; \citealt{Kormendy13}; \citealt{Heckman14}).

AGN-driven energetic outflows are a promising means by which an AGN can
interact with the gas in their larger-scale environment. Observations
of AGN, across a wide range of redshifts, have now revealed outflows in atomic, molecular and ionized
gas that are often identified over kiloparsec scales (e.g.,
\citealt{Veilleux05}; \citealt{Nesvadba06,Nesvadba08};
\citealt{Alexander10}; \citealt{Feruglio10};
\citealt{Lehnert11}; \citealt{Greene11}; \citealt{Alatalo11};
\citealt{Harrison12a, Harrison14b};
\citealt{Liu13b}; \citealt{Rupke13}; \citealt{Veilleux13}; \citealt{Genzel14};
\citealt{Arribas14}; \citealt{Brusa14}). A variety
of processes have been attributed to driving such outflows, including
radio jets, stellar winds, supernovae and quasar disk winds (e.g., \citealt{Holt08}; \citealt{Lehnert11}; \citealt{Guillard12};
\citealt{Rupke13}; \citealt{Zakamska14}). Some of the most powerful
analyses to address this issue come from combining spatially-resolved,
multi-wavelength data sets. In particular, in several AGN with high
radio-to-optical luminosity ratios, i.e., ``radio-loud'' AGN,
radio jets appear to be spatially-coincident with outflowing material
(e.g., \citealt{vanBreugel85}; \citealt{VillarMartin99a};
\citealt{Oosterloo00}; \citealt{SolorzanoInarrea03}; \citealt{Morganti05b,Morganti13}; \citealt{Stockton07};
\citealt{Nesvadba08}; \citealt{Shih13}; \citealt{Mahony13}). However, most AGN are
{\em radio quiet} (e.g., $\approx$85--90\% for optically-identified
quasars; e.g., \citealt{Kellermann89}; \citealt{Zakamska04}; \citealt{Lal10}) and, crucially, for these sources the
driver of the observed outflows remains less clear. 

To assess the prevalence, drivers and properties of ionized outflows in low-redshift luminous AGN, we 
performed multi-component fitting to the emission lines of
$\approx$24,000 $z<0.4$ optically identified AGN, which are predominately radio quiet
(\citealt{Mullaney13}). In this study we found that there is a relationship between
the presence of the most extreme ionized gas kinematics and radio luminosity. More specifically,
AGN with moderate-to-high radio luminosities (i.e., $L_{{\rm 1.4\,GHz}}\gtrsim10^{23}$\,W\,Hz$^{-1}$) are much more
likely to show broad and high-velocity kinematic components in their [O~{\sc
  iii}]$\lambda5007$ emission-line profiles than AGN with lower radio
luminosities.\footnote{The [O~{\sc
  iii}]$\lambda5007$ emission line traces warm ($T\sim10^{4}$\,K) ionized gas. As a
forbidden line, it cannot be produced in the high-density AGN broad-line region and is a good tracer of the low
density gas in galaxies over parsecs to tens of kiloparsecs scales (e.g.,
\citealt{Wampler75}; \citealt{Osterbrock89}).} A similar
result was reached by \cite{Zakamska14} for 568, $z<0.8$,
type~2 radio-quiet quasars and is also consistent with the results of several
other studies with smaller samples of AGN (e.g., \citealt{Heckman81}; \citealt{Wilson85};
\citealt{Veilleux91b}; \citealt{Whittle92}; \citealt{Nelson96};
\citealt{Nesvadba11}; \citealt{Husemann13}; \citealt{VillarMartin14}). However, there is an ongoing investigation into the origin of the
radio emission in radio-quiet AGN and therefore, it is not yet clear what
physical processes are responsible for the observed correlation between
ionized gas kinematics and radio luminosity (e.g., \citealt{Mullaney13}; \citealt{Zakamska14}). Overall, there are three main possibilities: (1) the radio
emission is dominated by star formation (e.g., \citealt{Sopp91}; \citealt{Condon13}) and associated stellar winds or
supernovae drive the outflows; (2) the presence of compact (i.e.,
$\lesssim$5\,kpc) radio jets/lobes dominate the radio emission
(e.g., \citealt{Kukula98}; \citealt{Ulvestad05}; \citealt{Singh14}) and also drive the outflows (e.g., \citealt{Whittle92}; \citealt{Leipski06b};
\citealt{Mullaney13}; \citealt{Kim13}); (3) the radio emission is dominated by nuclear coronal activity (e.g.,
\citealt{Laor08}) and/or shocks (\citealt{Zakamska14}; \citealt{Nims14}) and quasar winds are responsible
for driving the outflows (e.g., \citealt{FaucherGiguere12b};
\citealt{Zubovas12}). We now aim to address the outstanding issues of
the origin of the radio
emission in radio-quiet AGN and which processes drive their observed outflows, by combining high-resolution radio data with spatially-resolved
spectroscopy on a well defined sample of optically luminous AGN (\citealt{Mullaney13}; \citealt{Harrison14b}).

In this pilot study, we combine high-resolution radio imaging and spatially-resolved spectroscopy for one radio-quiet type~2 quasar, J1430+1339
(nicknamed the ``Teacup AGN''; \citealt{Keel12}; see
Section~\ref{Sec:Teacup}). This source is drawn from a larger sample of
type~2 AGN that host kiloparsec-scale ionized outflows (\citealt{Harrison14b}). We present new Karl G.\,Jansky Very Large Array (VLA) continuum
observations, new VIsible MultiObject Spectrograph
(VIMOS) integral field unit (IFU) observations and archival {\em Hubble
Space Telescope} ({\em HST}) narrow-band and medium-band
images. In Section~\ref{Sec:Teacup} we give background information on this
source; in Section~\ref{Sec:Data} we present the details of the data acquisition and reduction; in
Section~\ref{Sec:Discussion} we discuss the observations and their
implications; and in Section~\ref{Sec:Conclusions} we
give our conclusions. In our analysis we use $H_{0}=71\kms$, $\Omega_{\rm{M}}=0.27$, 
$\Omega_{\Lambda}=0.73$. Throughout, we relate flux density, $S_{\nu}$,
frequency, $\nu$, and radio spectral index, $\alpha$, in the
form $S_{\nu}\propto\nu^{\alpha}$. Images are displayed with north up
and east left and positional angles (PAs) are given east of north. 


\section{J1430+1339: The ``Teacup'' AGN}
\label{Sec:Teacup}

This study focuses on the radio-quiet type~2 quasar
SDSS\,J143029.88+133912.0 (J1430+1339) with a
redshift of $z=0.08506$. Using the Sloan Digital Sky Survey
(SDSS) spectrum, \cite{Reyes08} and
\cite{Mullaney13} independently identified this source as a type~2
(`obscured') AGN. The high [O~{\sc iii}] luminosity of $L_{{\rm [O
    ~III]}}=5\times10^{42}$\,erg\,s$^{-1}$ and the bolometric
AGN luminosity of $L_{{\rm  AGN}}=2\times10^{45}$\,erg\,s$^{-1}$ (derived
from the mid--far-infrared spectral energy distribution [SED]),
classifies this source as a type~2 {\em quasar} (\citealt{Reyes08};
\citealt{Harrison14b}; also see \citealt{Gagne14}). The moderate radio
luminosity of J1430+1339 ($L_{{\rm 1.4\,GHz}}=5\times10^{23}$\,W\,Hz$^{-1}$),\footnote{The radio
  luminosity of $L_{{\rm 1.4\,GHz}}=5\times10^{23}$\,W\,Hz$^{-1}$ is derived using the
flux density measurement of $S_{\rm 1.4\,GHz}=26.4\pm 0.4$\,mJy in
the $\approx$5\,arcsec resolution VLA FIRST imaging by \cite{Becker95}, and
assuming a spectral index of $\alpha=-1$.} results in this type~2 quasar being classified as {\em radio
quiet} in the $\nu L_{\nu}$(1.4\,GHz)-$L_{{\rm [O~III]}}$ plane (following
\citealt{Xu99}; \citealt{Zakamska04}). Additionally, the far-infrared
(FIR; 8--1000$\mu$m) luminosity attributed to star formation is unexceptional, with
$L_{{\rm IR,SF}}=3\times10^{44}$\,erg\,s$^{-1}$, which corresponds
to a star formation rate (SFR) of $\approx7$\,\msun\,yr$^{-1}$ using the
standard conversion from \cite{Kennicutt98} but converted to a
Chabrier initial mass function (see details in
\citealt{Harrison14b}).  Based upon the radio and infra-red luminosities,
J1430+1339 appears to be representative of low redshift quasars. However, as noted by \cite{Harrison14b} and \cite{VillarMartin14}, this source lies
more than a factor of 10 above the radio--far-infrared
correlation (following e.g., \citealt{Helou85}; \citealt{Ivison10b}), making this a {\em radio
  excess} source, with a radio excess parameter of $q_{{\rm
    IR}}=1.3\pm0.2$ (\citealt{Harrison14b}). This implies that processes
in addition to star formation are producing the radio emission
(such as radio jets, nuclear coronal activity or shocks; see
\citealt{Harrison14b}). In this paper we present new high-resolution
radio continuum observations to assess the processes that are producing the radio emission in J1430+1339. 

The properties of the ionized gas in J1430+1339 have resulted in
recent interest in the source. Using a citizen science campaign to
search for candidate AGN-ionized extended emission-line regions (EELRs) in SDSS images,
participants of the ``Galaxy Zoo'' project (\citealt{Lintott08})
identified J1430+1339 as having an interesting EELR morphology
(\citealt{Keel12}; later confirmed by {\em HST} imaging; \citealt{Keel14}). Due to
the loop-shaped ``handle''  of emission-line gas extending
$\approx$12\,kpc to the north-east of the SDSS image,
\cite{Keel12} nicknamed this galaxy the ``Teacup'' AGN (which is the
name we will now use). The {\em HST} imaging also reveals that the Teacup AGN resides in a bulge-dominated galaxy, with shell-like features, indicative of previous merger activity
(\citealt{Keel14}). The same team recently presented longslit data
on the ``Teacup'' AGN, and used photoionization modelling to
demonstrate that the EELR is dominated by photoionization by a central
AGN. They suggest that the AGN has dropped
in luminosity by one-to-two orders of magnitude over the last
$\approx$50,000 years (\citealt{Gagne14}) . 

Using integral field unit (IFU) spectroscopy we have shown that the Teacup AGN hosts kiloparsec-scale
ionized outflows (\citealt{Harrison14b}). Based upon the catalog of
\cite{Mullaney13} we initially identified a broad [O~{\sc iii}]
emission-line profile in its SDSS spectrum, indicative of ionized
outflows (this was also noted independently by \citealt{VillarMartin14}) and then
performed observations with Gemini's Multi-Object Spectrograph (GMOS)
IFU, in order to spatially resolve the ionized gas kinematics. These IFU
observations revealed a high-velocity kinematic component with $v\approx
-800$\,km\,s$^{-1}$, relative to the local gas velocity, located $\approx$\,1\,kpc
from the nucleus. Due to the small field of view of the GMOS
IFU used (i.e., $3.5\times5$\,arcsec), the full extent of the emission-line
region was not covered by these observations (including the large
emission-line loop described above). In this paper we
present IFU observations that cover the full $\approx20\times$15\,arcsec extent of the EELR, which enables us to trace the full kinematic structure of the
ionized gas.


 \section{Data acquisition, reduction and analysis}
\label{Sec:Data} 

For this paper we combine multiple datasets on the Teacup
AGN: (1) new radio imaging data obtained with the VLA
(Section~\ref{Sec:RadioData}); (2) new optical IFU data obtained
with ESO-VIMOS (Section~\ref{Sec:IFUdata}) and (3) archival {\em HST}
imaging (Section~\ref{Sec:HSTdata}). In this section we give
details of the observations, the steps of data reduction and the
analyses that we performed on each data set. In
Section~\ref{Sec:Discussion} we discuss the results using the combined
data sets.


\subsection{VLA data}
\label{Sec:RadioData}

\begin{figure} 
\centerline{\includegraphics[angle=90, scale=0.80]{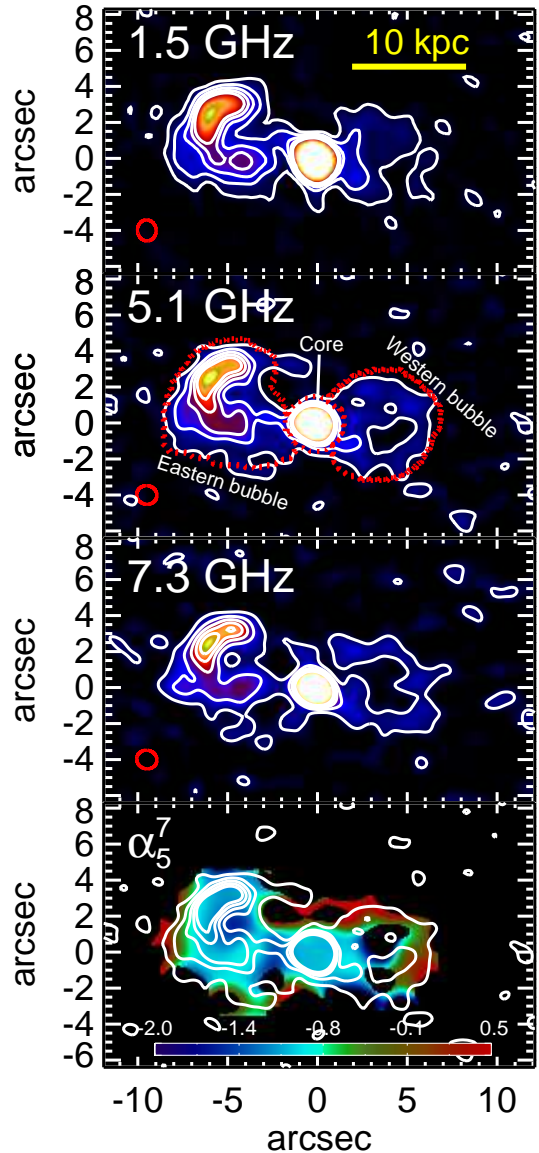} }
\caption{Our VLA data for the Teacup AGN at three different
  observed-frame frequencies (from top-down: 1.52\,GHz; 5.12\,GHz;
  7.26\,GHz), which have been matched to the same spatial resolution (see
  Section~\ref{Sec:RadioAnalysis}). The bottom panel shows a spectral
  index map using the full $C$-band data ($\alpha_{5}^{7}$), where values are only shown for pixels with an
  uncertainty $\le0.25$. The overlaid contours in this bottom panel are from the
  5.12\,GHz map. The contours in each panel indicate levels
  of [2, 6, 10, 14, 18]$\sigma$. The red ellipses represent the beams for each
  map and the yellow bar in the top panel represents 10\,kpc in
  length. We observe three distinct spatial structures: a $\approx$12\,kpc eastern bubble, a $\approx$11\,kpc western bubble and a
bright core. The red dotted contours overlaid on the 5.12\,GHz image
define the regions that we used for calculating the flux densities and
sizes of these structures. The regions of bright emission show steep
spectral indices of $\alpha^{7}_{5}\approx-1$. The SED for each of
these structure is shown in Figure~\ref{fig:RadioSED}.} 
\label{fig:RadioBubbles} 
\end{figure} 

\subsubsection{Observations}
We observed the Teacup AGN between December 2013 and May
2014 using the VLA at $L-$ and $C-$Band ($\approx1$--2\,GHz and $\approx4$--$8$\,GHz
respectively), in both the B$-$ and A$-$configurations. These
observations were taken under ID 13B-127, and were scheduled
dynamically to ensure appropriate weather conditions. We
bracketed our scans of the Teacup with 2\,minute scans of the bright, nearby BL Lac object J1415+1320, which we use for
phase referencing. We calibrated both the bandpass and the absolute flux
scale using scans of J1331+3030 (3C\,286) taken at the
beginning and the end of each session. At $L$-Band, we recorded 16 contiguous spectral windows in full polarization of $64\times1$\,MHz channels each,
that yields a total instantaneous bandwidth of 1024\,MHz. Accounting for losses
due to flagging of radio frequency interference (RFI), the band center
is 1.52\,GHz. At $C$-Band we recorded data in two frequency bands of eight contiguous spectral windows
in full polarization, centered on $\approx$5\,GHz and $\approx$7\,GHz, with each spectral window consisting of $64\times2$\,MHz channels, yielding $1024\times2$\,MHz channels in
total. The raw visibilities of the VLA data were calibrated locally using Version 1.2.0 of
the {\sc casa} VLA Calibration pipeline.\footnote{https://science.nrao.edu/facilities/vla/data-processing/}

\subsubsection{Radio maps and analysis}
\label{Sec:RadioAnalysis}

\begin{table}
\begin{center}
{\footnotesize
\caption{Properties of Radio Images}
\begin{tabular}{cccccccc}
\hline
Obs. Frequency   & Beam HPBW      & Beam PA       & Noise & Figure \\ 
(GHz)                 & (arcsec)    & (degrees) & ($\mu$Jy) &  \\ 
\noalign{\smallskip}
\hline
1.52                & 1.17$\times$1.04  &$-172$  & 38& \ref{fig:RadioBubbles}  \\
5.12                & 1.13$\times$1.04  & 80         & 16& \ref{fig:RadioBubbles}   \\
7.26                & 1.14$\times$1.03  & 74         & 15&  \ref{fig:RadioBubbles}   \\
6.22                & 0.51$\times$0.38  & $-61$ & 10 &  \ref{fig:RadioHighRes}   \\
6.22                & 0.37$\times$0.24  & $-88$ & 18 & \ref{fig:RadioHighRes}(a) \\
5.12                & 0.50$\times$0.30  & 87       & 28 & \ref{fig:RadioUHRes} \\
7.26                & 0.36$\times$0.21  & 86       & 27 & \ref{fig:RadioUHRes} \\
\hline
\hline
\label{Tab:Images}
\end{tabular}
}
\tablecomments{A summary of the central observed frequencies, synthesized beams and
  noise of the main radio maps used in this work. In
  Section~\ref{Sec:RadioAnalysis} we describe which sub-sets of the
  VLA data went into each map and how they were produced. The final column
  gives the figure numbers where the maps are shown.}

\end{center}
\end{table}

Below we describe a number of maps that we created from our data at different resolutions and with
different frequency centroids. The details of the synthesized beams
and the noise in each map are summarized in
Table~\ref{Tab:Images}. We also provide details of how we calculated
flux densities, their uncertainties and spectral indices for the
radio structures that we identified. These values are summarized in Table~\ref{Tab:RadioFluxes}.

In the top panel of Figure~\ref{fig:RadioBubbles} we show the $L$-Band,
A-configuration VLA data (band center
is 1.52\,GHz), mapped with a {\sc robust}=0.5 weighting
scheme and no additional tapering in the $uv$ plane. We achieve a
synthesized beam of $\approx$1\,arcsec. In the second and third panels of Figure~\ref{fig:RadioBubbles}, we
show maps from two sub-sets of the $C$-band data, centered at 5.12\,GHz
and 7.26\,GHz that are mapped by weighting the $uv$ data with Gaussian tapers of {\sc
  fwhm}=$120$\,k$\lambda$ and $100$\,k$\lambda$, respectively, to
achieve a close approximation with the resolution of the $L$-band
imaging (see Table~\ref{Tab:Images}). In the bottom panel of Figure~\ref{fig:RadioBubbles}  we show
a radio spectral index map (between 7\,GHz and 5\,GHz; $\alpha^{7}_{5}$) from the full $C$-band
spectral data set that was mapped using the multi-scale, multi-frequency
synthesis ({\sc msmfs}) imaging mode in {\sc casa}. The resolution of this spectral index map is
$2.05\times 1.78$\,arcsec (${\rm PA}=74^{\circ}$).

\begin{figure} 
\centerline{\includegraphics[angle=90, scale=0.35]{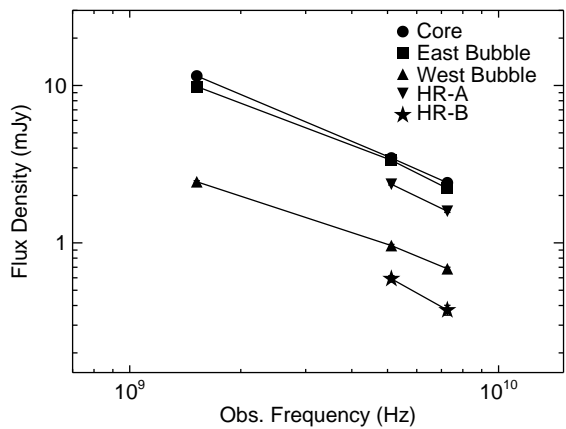} }
\caption{The VLA data for each of the spatial structures identified
in Figure~\ref{fig:RadioBubbles} and Figure~\ref{fig:RadioHighRes}. We show a SED for each of these
structures, where we have fit a piecewise power-law between each data
point. The error bars are plotted but are generally smaller than
the symbol sizes. These data are tabulated in
Table~\ref{Tab:RadioFluxes}. While all of the features are dominated
by steep spectral indices (i.e., $\alpha\approx-1$) we note that there
is some variation within individual regions (see Figure.~\ref{fig:RadioBubbles}).} 
\label{fig:RadioSED} 
\end{figure} 

From the three matched-resolution maps described above (Figure~\ref{fig:RadioBubbles}), we identify three spatial
structures: a bright eastern bubble, a fainter western bubble, and a
bright core. We measured the flux densities at 1.52\,GHz, 5.12\,GHz and
7.26\,GHz of these features, from the maps shown in
Figure~\ref{fig:RadioBubbles}, by integrating in polygons around each feature (with the edges of the
polygons motivated by the $2\sigma$ contours in the 5.12\,GHz
map; see Figure~\ref{fig:RadioBubbles}). The measured flux densities are reported in
Table~\ref{Tab:RadioFluxes}, where the uncertainties correspond to the
noise in the respective maps. Using our three-band radio photometry we fit piecewise
power-law SEDs between $1.52$--$5.12$\,GHz and between
$5.12$--$7.26$\,GHz (Figure~\ref{fig:RadioSED}) and calculate the
spectral indices ($\alpha^{5}_{1.5}$ and
$\alpha^{7}_{5}$; Table~\ref{Tab:RadioFluxes}). To reflect the
variation of the spectral indices within each region, in
Table~\ref{Tab:RadioFluxes}, we also quote the
1$\sigma$ scatter, within these regions, derived from the spectral
index map shown in Figure~\ref{fig:RadioBubbles}. We note that the
total $1.52$\,GHz flux density across all regions, seen in this
$\approx$1\,arcsec resolution $L$-band map, is $S_{\rm
  1.5\,GHz}=23.79\pm0.07$\,mJy; however, by re-imaging our $L$-band data
to a resolution $4.59\times4.18$\,arcsec we recovered a total of $S_{\rm
  1.5\,GHz}=25.9\pm 0.1$\,mJy. We thus conclude that $\approx$2\,mJy
of the $L$-band radio flux in the Teacup AGN is produced in a diffuse, low surface brightness halo surrounding the core and bubbles, whose emission is resolved-out
in our higher-resolution continuum maps. 

\begin{table*}
\begin{center}
{\footnotesize
\caption{Radio Properties of Selected Structures}
\begin{tabular}{lcccccccc}
\hline
Structure & $D$                       & {\sc PA} & $S_{{\rm
    1.5\,GHz}}$ & $S_{{\rm 5\,GHz}}$ & $S_{{\rm 7\,GHz}}$ &
$\alpha^{5}_{1.5}$ &$\alpha^{7}_{5}$&Variation in $\alpha$ \\
               & (1)             &   (2)    & (3)                    & (4)                 &  (5)                &         (6)                   &       (7)   & (8)             \\
\noalign{\smallskip}
\hline
\multicolumn{9}{c}{Matched Resolution (Figure~\ref{fig:RadioBubbles})} \\
Core                  &1.5[2]    &0   &11.49[4]& 3.462[16]   & 2.410[15]  & -0.988[5] & -1.04[2] & $\pm$0.08     \\
Eastern bubble  &7.9[5]    &66    &9.85[4]  & 3.350[16]   & 2.219[15]  & -0.888[5] & -1.18[2] & $\pm$0.3      \\
Western bubble &6.9[5]    &$-90$ &2.45[4]  & 0.961[16]   & 0.685[15]  & -0.77[2] & -0.98[8] & $\pm$0.6   \\
\hline
\multicolumn{9}{c}{High Resolution (Figure~\ref{fig:RadioHighRes} and Figure~\ref{fig:RadioUHRes})}\\
HR-A            &       $\lesssim0.4$                 &- &    -       &  2.36[3]     & 1.59[3]      &  -  &    -1.13[6] & -\\
HR-B            &        $\lesssim0.4$                &-  &   -      &   0.59[3]     & 0.37[3]      &  -   &    -1.3[2]     &- \\
\hline
\hline
\label{Tab:RadioFluxes}
\end{tabular}
}
\tablecomments{The sizes, flux densities and spectral indices of the
  structures identified in the VLA data of the
Teacup AGN (see Figure~\ref{fig:RadioBubbles} and
Figure~\ref{fig:RadioHighRes}). (1) The projected size of each
structure (in arc seconds) based on the 5\,GHz images, measured by taking a vector from
the central core peak to the edge of each region (in the case of HR-A
and HR-B we quote the HPBW); (2) The {\sc PA} (in degrees) of the axis
that the distances, $D$, were measured (for the eastern bubble this is in the direction to the
brightest knot); (3)--(5) The observed-frame flux densities in mJy; (6) \& (7) The spectral indices
calculated from the quoted flux densities; (8) The scatter of the $\alpha$ values within the defined
regions (see Figure~\ref{fig:RadioBubbles}). In each column the number in square brackets gives
the uncertainty on the last decimal place. We estimate an additional
$\approx$5\% ($\approx$15\%) uncertainty on the flux densities
of the core and eastern bubble (western bubble) due to the uncertain
definitions of the chosen regions (Figure~\ref{fig:RadioBubbles}).}

\end{center}
\end{table*}

\begin{figure}
\centerline{\includegraphics[angle=90, scale=0.34]{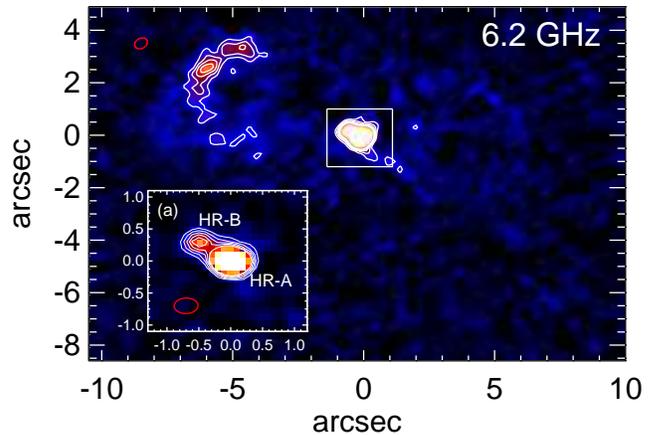} }
\caption{Our high-resolution VLA data for the Teacup AGN at
  observed-frame 6.22\,GHz. The
  eastern bubble and bright core can clearly be observed, whilst some
  of the fainter extended emission has been resolved out (see Figure~\ref{fig:RadioBubbles}). Inset (a)
  shows our highest resolution map (see Table~\ref{Tab:Images}), zoomed-in around the
  white box indicated in the main panel. The contours in both maps are
  at levels of [5,8,11,...]$\sigma$. The red ellipses represent
  the beams. These data clearly show that the
  bright core, observed in Figure~\ref{fig:RadioBubbles}, is composed of
  two spatial structures: a bright central structure
  (HR-A) and a fainter structure (HR-B) $\approx$0.8\,kpc to the north east
  (${\rm PA}\approx60^{\circ}$). The HR-B
  structure is very closely aligned in PA with the brightest region on the edge of the eastern
  bubble.}
\label{fig:RadioHighRes} 
\end{figure} 

\begin{figure}
\centerline{\includegraphics[angle=0, scale=0.5]{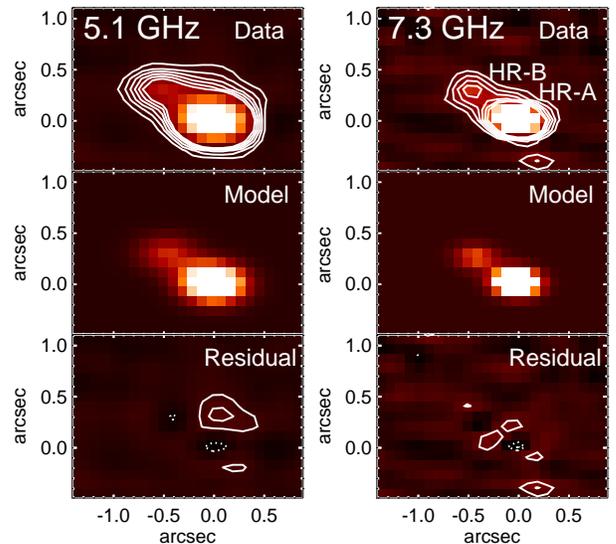} }
\caption{Our highest resolution ($\approx$0.4\,arcsec; see Table~\ref{Tab:Images}) VLA data at 5.12\,GHz and
  7.26\,GHz for the central few arc seconds of the Teacup AGN. We have
  labeled the two unresolved structures HR-A and HR-B, which we identified in
  Figure~\ref{fig:RadioHighRes}. The top row shows the data, the second
  row shows our fits of two beams to these data (``model'') and the bottom row shows
  the data minus the fit (``residual''; see Section~\ref{Sec:RadioAnalysis}). In each panel
  the contours represent levels of
  [...,$-7$,$-5$,$-3$,3,5,7,...]$\sigma$, where the negative contours are
  dotted lines and positive contours are solid lines. The majority of
  the flux is inside two unresolved beams (corresponding to HR-A and
  HR-B) with some low-significance emission between these two structures. The flux
  densities and spectral indices ($\alpha_{5}^{7}$) for these two structures
  are provided in Table~\ref{Tab:RadioFluxes}.}
\label{fig:RadioUHRes} 
\end{figure}

To search for spatial structures on scales smaller than
$\approx$1\,arcsec, we make use of the $C$-band data at full
resolution. In Figure~\ref{fig:RadioHighRes} we show a {\sc
  robust}=0.5 weighted map, created from the B- and A-configuration,
across the entire $C$-band, centered at 6.22\,GHz. In this $\approx$0.5\,arcsec resolution map the fainter western bubble is resolved out, while the
core appears to be slightly elongated along a position angle of
PA$\approx$60$^\circ$; i.e., in line with the position angle connecting the core to the
brightest region in the eastern bubble (see
Table~\ref{Tab:RadioFluxes}). Inset\,(a) of
Figure~\ref{fig:RadioHighRes} shows our highest, $\approx0.3$\,arcsec, resolution $C$-band map of the core, comprising only of the
A-configuration $uv$ data that is mapped with super-uniform
weighting in {\sc casa}. In this map, the core is clearly resolved
into two structures, which we label ``High Resolution'' (HR)-A and
HR-B. HR-B, with peak position: 14$^{\rm h}$30$^{\rm m}$29$^{\rm s}$.902, +13$^{\circ}$39$^{\prime}$12$^{\prime\prime}$.15, is located
0.5\,arcsec ($\approx$0.8\,kpc) to the
north-east, at ${\rm PA}\approx60^{\circ}$, of HR-A which has peak position: 14$^{\rm h}$30$^{\rm m}$29$^{\rm s}$.873,
+13$^{\circ}$39$^{\prime}$11$^{\prime\prime}$.90. To measure the 5\,GHz and 7\,GHz flux densities of these
two structures we fit super-uniform weighted 5\,GHz and 7\,GHz maps
with two beams (i.e., two unresolved components;
Figure~\ref{fig:RadioUHRes}). The majority of the flux in
the core is indeed inside two unresolved structures (corresponding to HR-A
and HR-B) at both 5\,GHz and 7\,GHz. Therefore, these structures are unresolved at
a half-power beam width of HPBW$\approx$0.4\,arcsec (i.e., $\approx$0.6\,kpc), which can be
taken as very conservative upper limits on their intrinsic
sizes. Figure~\ref{fig:RadioUHRes} reveals low surface brightness emission (at $\approx3\sigma$) between the
structures in the residual image. Furthermore, the combined 5\,GHz flux density of HR-A and HR-B
($S_{{\rm 5GHz}}=2.95$\,mJy; Table~\ref{Tab:RadioFluxes}) is less than
that of the total core structure from the lower resolution map ($S_{{\rm 5GHz}}=3.46$\,mJy), implying a further
$\approx$0.5\,mJy of diffuse material the may exist inside the core that
is not part of these two dominant structures.

\subsection{VIMOS IFU data}
\label{Sec:IFUdata}

\subsubsection{VIMOS IFU data: observations}
We observed the Teacup AGN with the VIMOS (\citealt{LeFevre03}) instrument installed on the ESO/VLT
telescope  between 2014 March 9 and 2014 March
10 (Program ID: 092.B-0062). The seeing was $<0.9$\,arcsec throughout. The observations were carried out in IFU mode using the
0.67 arcsec fibre, which provides a field-of-view of
27$\times$27\,arcsec, and the HR-Orange grism,
which provides a wavelength range of 5250\AA--7400\AA\ at a spectral
resolution of $\approx$2650. By using the widths of sky-lines from our
observations we found that observed resolution was
${\rm FWHM}=105\pm$8\,km\,s$^{-1}$ around $\lambda\approx5400$\,\AA
(i.e., around the observed wavelength of [O~{\sc iii}]$\lambda$5007). We corrected our measurements for this instrumental resolution.  During the observations
the target was dithered around the four quadrants of the VIMOS
IFU. There were nine on-source exposures performed, each of 540 seconds,
resulting in a total on-source exposure time of 4860s. The on-source
exposures were interspersed with three on-sky exposures, that were used
for sky-subtraction. Flat field frames
and arc-lamp frames were observed periodically during the
observations. A standard star, taken under similar conditions to the
science observations was also observed.

The data reduction on the VIMOS data was performed using the standard
{\sc esorex} pipeline, which
includes bias subtraction, flat-fielding and wavelength
calibration. The standard star was reduced in an identical manner to the science frames and {\sc esorex}
was used to apply the flux calibration. Data cubes
were constructed from the individually sky-subtracted, reduced science frames and
consequently the final data cube was created by median combining these
cubes using a three-sigma clipping threshold.  

\subsubsection{VIMOS IFU data: emission-line fitting}
\begin{figure} 
\centerline{\includegraphics[angle=0, scale=0.95]{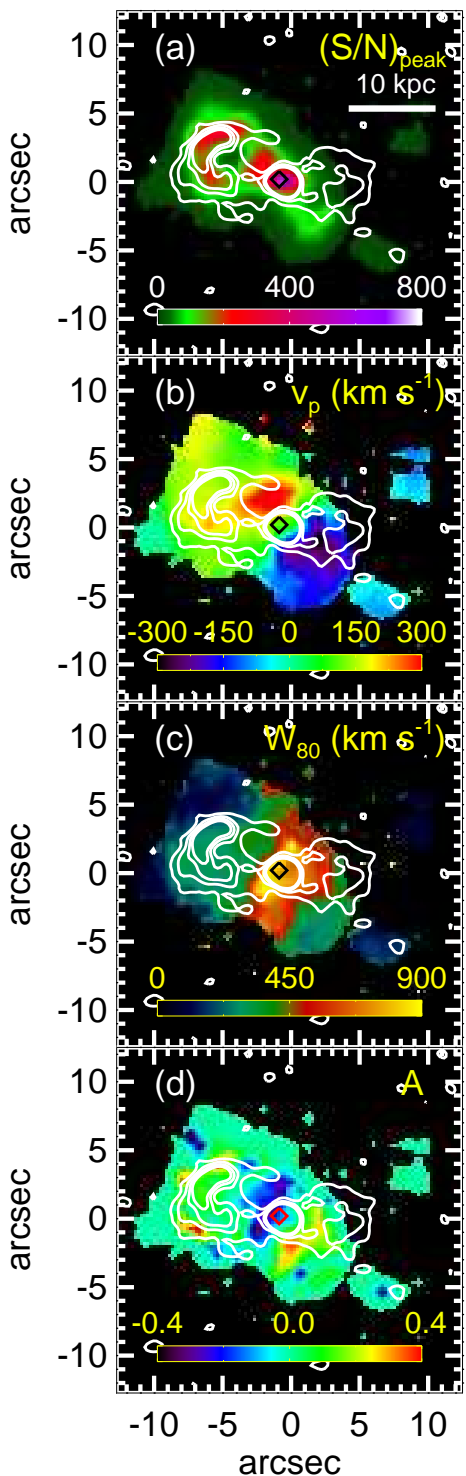}}
\caption{Our VIMOS IFU data for the Teacup AGN. These maps characterize the
  [O~{\sc iii}] emission-line profiles, at each spatial pixel, using the non-parametric
  definitions described in Section~\ref{Sec:IFUdata}. From
  top-to-bottom: (a) The signal-to-noise of the peak flux density; (b) the velocity
  of the peak flux density; (c) the
  emission-line width; (d) the asymmetry of the emission-line profile. The
  contours in each panel show the 5\,GHz radio data from
  Figure~\ref{fig:RadioBubbles}. The diamond in each panel shows the
  position of the HR-B radio structure seen in
  Figure~\ref{fig:RadioHighRes}. The dominant velocity gradient seen
  in panel (b) appears to be independent of the distribution of radio emission; however, there is bright ionized gas associated
with the luminous eastern radio bubble seen in panel (a). Panels (c)
and (d) reveal very broad and asymmetric emission-line profiles around the position of the HR-B
radio structure, this is due to an additional high-velocity kinematic
component at this location (see Figure~\ref{fig:HST}).} 
\label{fig:VelMaps} 
\end{figure} 

Our IFU data for the Teacup AGN covers $\approx$40$\times$40\,kpc,
which means that we can trace the kinematic structure over the full extent of
the EELR. The analysis performed on the IFU data in this work closely follows
the methods described in detail in Section~3 of \cite{Harrison14b};
therefore, we only give brief details here. 

To account for the complexity in the emission-line profiles we opted
to use non-parametric definitions to characterize the emission-line
profiles within each pixel of the IFU data. In this work we use the
following definitions:
\begin{enumerate}
\item The peak signal-to-noise, (S/N)$_{{\rm peak}}$, which is the
  signal-to-noise of the emission-line profile at the peak flux density. This
  allows us to identify the spatial distribution of the emission-line
  gas, including low surface-brightness features. 
\item The ``peak velocity'' ($v_{\rm p}$), which is the velocity of
  the emission-line profile at the peak flux density. For the Teacup AGN,
  this traces the velocity structure of the dominant narrow component
  of the emission-line profile (see \citealt{Harrison14b}).
\item The line width, $W_{80}$, which is the velocity that
  contains 80\% of the emission-line flux. This is equivalent to 1.088$\times${\sc
    FWHM} for a single Gaussian profile. 
\item The asymmetry value ($A$; see \citealt{Liu13b}), which is
  defined as 
\begin{equation}
A \equiv \frac{(v_{90}-v_{50})-(v_{50}-v_{10})} {W_{80}}, 
\end{equation}
where $v_{10}$, $v_{50}$, $v_{90}$, are the velocities of at 10th,
50th and 90th percent of the cumulative flux, respectively. A very
negative value of $A$ means that the emission-line profile has a strong
blue wing, while $A=0$ indicates that it is symmetric. 
\end{enumerate}

To illustrate the kinematic structure of the ionized gas, using our IFU data, we
use the [O~{\sc iii}]$\lambda$5007 emission line as it is the
brightest line in the spectra and it is not blended with other emission lines. In Figure~\ref{fig:VelMaps} we show four
maps, produced using this [O~{\sc iii}] emission line, corresponding to the four non-parametric values described
above. The values were calculated at each pixel of the data cube following the methods of \cite{Harrison14b}. Briefly, we
first fit the emission-line profiles (weighting against wavelengths with bad sky lines) with multiple Gaussian
components before using the emission-line fits to define each of
these four values. By using the emission-line fits we are able to
define our non-parametric values even in regions of low
signal-to-noise (see \citealt{Harrison14b}). Values are assigned only to pixels where the
emission-line was detected with (S/N)$_{{\rm peak}}\ge5$. In
Figure~\ref{fig:VelMaps}, we re-sampled the pixel scale of the maps up by a factor
of three, solely for the purposes of a clearer comparison to the
morphology of the radio data.

We extracted spectra from different spatial regions of the IFU data cube,
where the regions were motivated by the morphology of the radio
emission. These spectra were obtained by summing the spectra from the pixels in the
vicinity of the three regions shown in Figure~\ref{fig:HST}: (1) the
eastern edge of the eastern bubble; (2) the western edge of the
western bubble; and (3) the region around the HR-B radio structure. We fit the
[O~{\sc iii}] emission-line profiles of these spectra in an identical
manner to that outlined above for the individual spatial
pixels. The fits to the spectra are shown in Figure~\ref{fig:HST}, and the
non-parametric values, derived from these fits are given in Table~\ref{Tab:Spectra}. We calculated uncertainties by fitting to 10$^{4}$ mock spectra
that were created by taking the best fit model and adding random
Gaussian noise. We found that these formal uncertainties were very small (i.e.,
$\lesssim$10\%) for the fits to
emission-line profiles of the eastern
bubble region and HR-B region and do not reflect the true
uncertainties, which are dominated by the range in values within each region. Therefore, in Table~\ref{Tab:Spectra} we provide the
range of each value across the individual pixels within the regions.

\begin{table}
\begin{center}
{\footnotesize
\caption{[O~{\sc iii}] emission-line profile properties from selected regions}
\begin{tabular}{lcccc}
\hline
Region & (S/N)$_{{\rm peak}}$ & $v_{{\rm p}}$   & $W_{80}$ & $A$ \\
           &           &  (km\,s$^{-1}$)     & (km\,s$^{-1}$)                    & \\
\noalign{\smallskip}
\hline
HR-B region              &640  &147$_{-40}^{+11}$  &920$_{-120}^{+50}$& -0.19$_{-0.15}^{+0.04}$  \\
East bubble edge  &805 &150$_{-40}^{+70}$  &290$_{-50}^{+20}$  & 0.14$_{-0.15}^{+0.02}$ \\
West bubble edge &5    &-110$\pm$20         &320$\pm$70  & 0($^{\star}$)  \\
\hline
\hline
\label{Tab:Spectra}
\end{tabular}
}
\tablecomments{The peak signal-to-noise (S/N)$_{{\rm peak}}$, peak velocity ($v_{{\rm
      p}}$), line-width ($W_{80}$) and
  asymmetry ($A$) values for the [O~{\sc iii}] emission-line
  profiles extracted from labelled regions in Figure~\ref{fig:HST}. The upper and lower limits on
  the values of HR-B and the eastern bubble, reflect the full range in
  values of the individual pixels inside the defined regions. As the western bubble is not
  detected in individual pixels (Figure~\ref{fig:VelMaps}), we provide
  the 2$\sigma$ uncertainties on the values for this region (see Section~\ref{Sec:IFUdata}).
\newline$^{\star}$ As only one Gaussian
  component is fit to the emission-line profile of the western bubble,
  $A=0$ by default (see Section~\ref{Sec:IFUdata}). 
}

\end{center}
\end{table}

\subsection{{\em HST} imaging}
\label{Sec:HSTdata}

\begin{figure*} 
\centerline{\includegraphics[angle=90, scale=0.8]{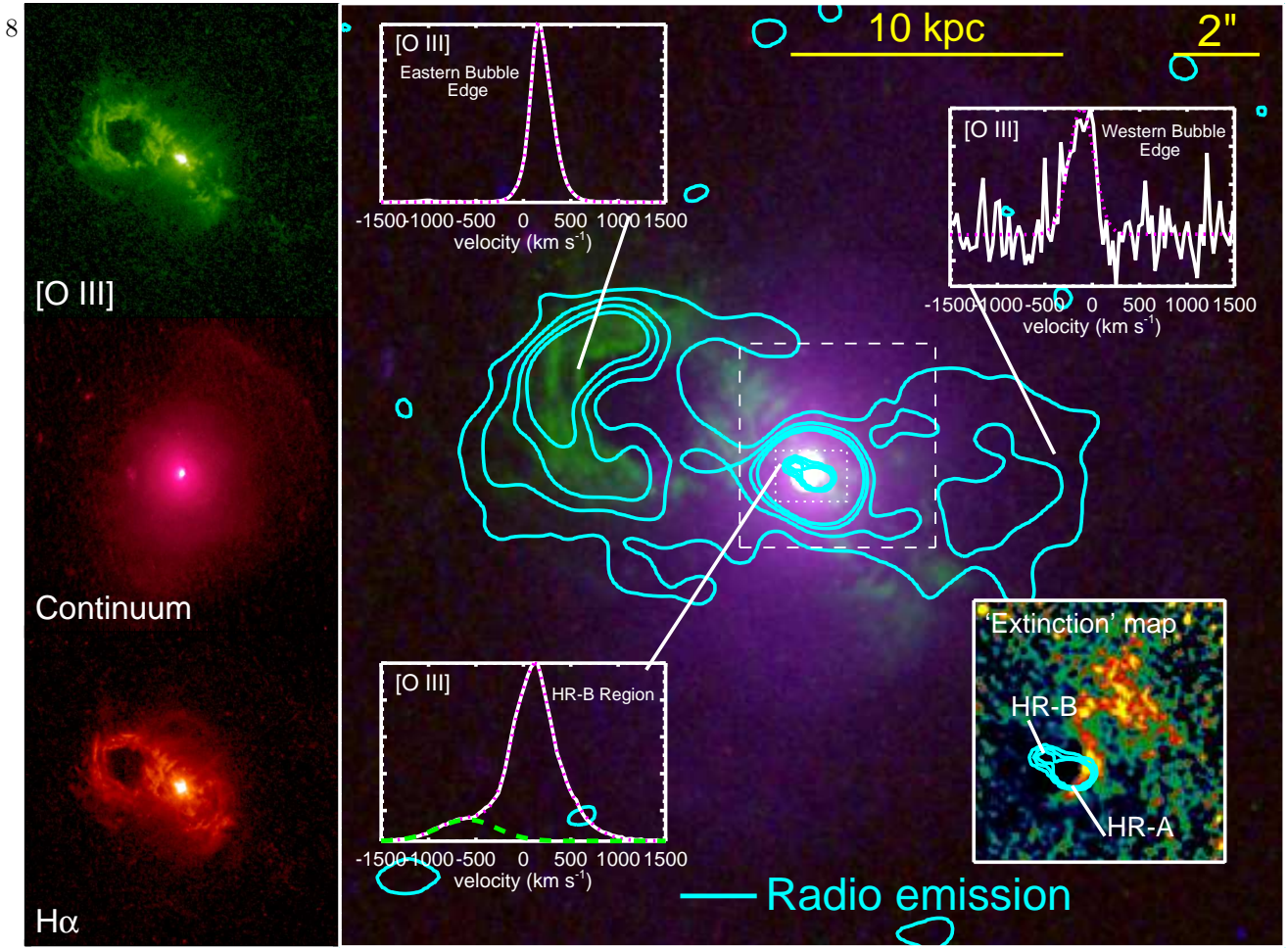}}
\caption{ {\em HST} images for the Teacup AGN (described in Section~\ref{Sec:HSTdata}). {\em Top Left:} Narrow-band image
  centered on [O~{\sc iii}]$\lambda$5007. {\em Middle Left:}
  Line-free continuum image. {\em Bottom Left:} Narrow-band image
  centered on H$\alpha$. {\em Right:} Composite image of the continuum
  image (shown in purple) and [O~{\sc iii}] emission (shown in green). The
  contours represent the 5.2\,GHz radio emission from
  Figure~\ref{fig:RadioBubbles}, except within the dotted white box where the
  contours represent the high-resolution 6.2\,GHz data from
  Figure~\ref{fig:RadioHighRes}. The three inset [O~{\sc
    iii}]$\lambda5007$ emission-line profiles are from our
  VIMOS IFU data cubes, extracted from the specified regions (see
  Section~\ref{Sec:IFUdata}). The fits to
the spectra are over plotted as magenta dotted lines and the values
derived from these fits are shown in Table~\ref{Tab:Spectra}. The
highest-velocity kinematic component ($v=-740$\,km\,s$^{-1}$; see Section~\ref{Sec:Description}) is overlaid as a dashed green
line on the ``HR-B Region'' emission-line profile. The
inset on the bottom-right is a pseudo-extinction
  map, extracted from the region indicated by the dashed white box
  (see Section~\ref{Sec:HSTdata}). Overlaid as contours
  are the high-resolution 6.2\,GHz radio data (see
  Figure~\ref{fig:RadioHighRes}). In this inset, the dusty regions can be observed, as
  brighter regions, extending from the core to $\approx$2\,arcsec north west of the
  nucleus. The radio structures are preferentially orientated away from
  the most dusty region.} 
\label{fig:HST} 
\end{figure*} 

The Teacup AGN has been observed with {\em HST}
and the data is published in \cite{Keel14}. We show the {\em HST} data in this
work to compare the emission-line and continuum morphologies to our
radio data and IFU data (see Figure~\ref{fig:HST}). Narrow-band images around
the [O~{\sc iii}]$\lambda$5007 and H$\alpha+$[N~{\sc ii}] emission lines
were taken using the FR551N and FR716N filters,
respectively, on the Advance Camera for Surveys (ACS) instrument. The continuum
emission, in the wavelength region of these two emission lines, was observed using the medium-band
filters F621M and F763M on the Wide-Field Camera 3 (WFC3)
instrument (the details of all of these {\em HST} observations can be
found in \citealt{Keel14}). We applied a small astrometric
correction to register the astrometry of the {\em HST} images with
the SDSS position of the Teacup AGN and our IFU data. In the left
column of Figure~\ref{fig:HST} we show the continuum-free
emission-line images around [O~{\sc iii}]$\lambda$5007 and
H$\alpha+$[N~{\sc ii}] and the average of the two line-free continuum
images, all on a logarithmic scale. In the main panel of Figure~\ref{fig:HST}
we also show a three-color composite image where red and blue are the two continuum
images (resulting in a purple continuum color) and green is the
[O~{\sc iii}] narrow-band image. Subtle features of dust lanes can be
seen in the {\em HST} continuum images, north-west of the core (see \citealt{Keel14}). To
illustrate where these dusty regions are, in the inset of Figure~\ref{fig:HST}, we
created a pseudo-extinction map where we have divided the two continuum
images. This map clearly shows a variation of optical colors in the
dustier regions identified by \cite{Keel14}, which is also confirmed by the
spatial variation of the H$\alpha$/H$\beta$ emission-line ratio
observed by \cite{Gagne14}. 


 
\section{Results and Discussion} 
\label{Sec:Discussion}
 
\begin{figure} 
\centerline{\includegraphics[angle=0, scale=0.6]{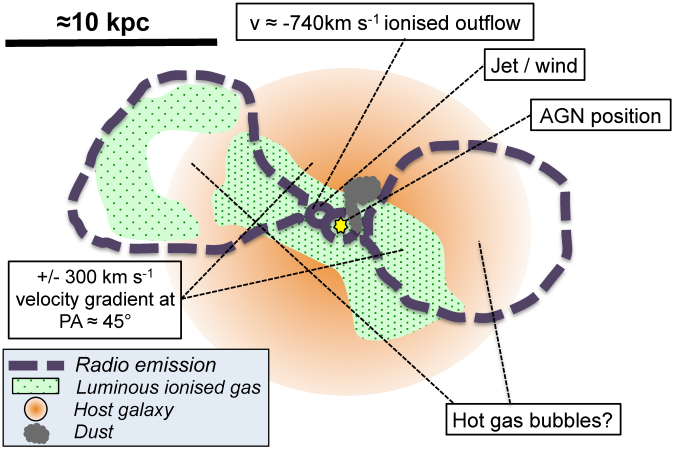}} 
\caption{A schematic diagram of the Teacup AGN to summarize the results of the data
 presented in this paper (see Section~\ref{Sec:Description}). We have
 highlighted the region of luminous ionized gas with dotted-green shading; however, we note that low surface brightness ionized
  gas is found over a much larger region (Figure~\ref{fig:VelMaps}). } 
\label{fig:Schematic} 
\end{figure}

In the previous section we presented a variety of data on the Teacup
AGN, a $z=0.085$ type~2 radio-quiet quasar (see
Section~\ref{Sec:Teacup} for background information), these were: high-resolution radio imaging at 1--8\,GHz
(Figure~\ref{fig:RadioBubbles} to Figure~\ref{fig:RadioUHRes}); optical IFU data
(Figure~\ref{fig:VelMaps}) and {\em HST} emission-line and continuum
images (Figure~\ref{fig:HST}). Our combined data sets have revealed a number of interesting kinematic and
morphological structures in the Teacup AGN and we have summarized the
most interesting features in a schematic diagram in
Figure~\ref{fig:Schematic}. In this pilot study, for future work on larger samples of radio-quiet AGN, the aim is to
understand the relationship between the ionized gas kinematics and the
radio emission from this source. In the following sub-sections we describe our main results and
discuss some possible interpretations of the observations.

\subsection{Summary of the source structure}
\label{Sec:Description}

Our VLA radio imaging of the Teacup AGN reveals two steep spectrum ($\alpha_{5}^{7}\approx-1$; see Table~\ref{Tab:RadioFluxes}) radio structures,
extending $\approx$10--12\,kpc each side of a central steep spectrum
core (see Figure~\ref{fig:RadioBubbles}). These structures appear
  to be bubbles filled with low surface brightness extended emission
  that have brighter edges. There is a flattening of the
spectral indices towards the edges of the bubbles, which is most
noticeable in the western bubble, where $\alpha_{5}^{7}\gtrsim-0.8$. Overall the total 1.5\,GHz radio emission is divided such that
$\approx$47\% of the emission is found in the two radio
bubbles, $\approx$44\% in a core region and $\approx$9\% is likely to
be in an extended diffuse structure (see Section~\ref{Sec:RadioAnalysis} and Table~\ref{Tab:RadioFluxes}). The
brighter eastern bubble is most luminous on the eastern edge, where it is spatially co-incident with an arc
of luminous ionized gas emission, seen in both [O~{\sc iii}] and H$\alpha$
(see Figure~\ref{fig:VelMaps} and Figure~\ref{fig:HST}). The {\em HST} imaging also reveals that this
gas has a filamentary structure that broadly follows the curvature of
the radio bubble (Figure~\ref{fig:HST}). The [O~{\sc iii}] and
  H$\alpha$ emission is much fainter at the location of the western
  radio bubble (Figure~\ref{fig:HST}). The gas may be misaligned with the ionizing
  radiation at this point or, alternatively, a low gas density could explain both the low radio
luminosity and lack of bright ionized gas emission at this position.

The {\em kinematic} structure of the ionized gas does not obviously
trace the morphology of the radio bubbles (see
Figure~\ref{fig:VelMaps}). The largest velocity gradient
(with $v_{p}\approx\pm$300\,km\,s$^{-1}$) is found within the host galaxy, extending to $\approx$5\,kpc each side of the
core with PA$\approx$45$^{\circ}$, i.e., mis-aligned with the
radio bubbles by $\approx$20$^{\circ}$. If this gas is a rotational
kinematic component, due to the host galaxy's gravitational potential, it implies a dynamical galaxy mass of
$\approx$10$^{11}$\,M$_{\odot}$. The gas at the edge of
the eastern bubble has a velocity offset of $v_{{\rm
    p}}=150$\,km\,s$^{-1}$ relative to the systemic, or
$\approx$$-150$\,km\,s$^{-1}$ relative to the possible
$\approx$300\,km\,s$^{-1}$ ``rotational'' component near this position. The gas at the edge of the western bubble has a velocity of
$v_{{\rm p}}=-110$\,km\,s$^{-1}$ or $\approx$190\,km\,s$^{-1}$
relative to the $\approx$$-300$\,km\,s$^{-1}$ ``rotational'' component. Therefore, the gas in the
regions around and beyond the radio bubbles may be decoupled from the
 possible ``rotational'' kinematic component. The emission-line profiles on the edges of both
bubbles have a moderate emission-line width of
$W_{80}\approx300$\,km\,s$^{-1}$, compared to those in the central
few kiloparsecs which are typically $\approx$500--900\,km\,s$^{-1}$
(see Figure~\ref{fig:VelMaps}). We discuss the possible origin of
the $\approx$10--12\,kpc radio bubbles and associated arc of ionized gas in Section~\ref{Sec:Bubbles}. 

Our highest resolution radio image of the Teacup AGN (Figure~\ref{fig:RadioHighRes}) reveals
that the bright central core, identified in
Figure~\ref{fig:RadioBubbles}, is dominated by two unresolved
structures (the resolution is HPBW$\approx$0.6\,kpc); a bright
central structure and a fainter structure $\approx$0.8\,kpc to the north east (${\rm PA}\approx60^{\circ}$). We named these
structures ``HR-A'' and ``HR-B'', respectively. Two intriguing points should be noted about the ``HR-B''
structure: (1) it is approximately along the same axis as the large eastern bubble (with
respect to the central core) and (2) it is co-located with a high-velocity
(blue-shifted) ionized gas component (see Figure~\ref{fig:HST}).  We note that this high-velocity kinematic
  component has also been observed with GMOS-IFU observations of this source
  (\citealt{Harrison14b} and W. Keel priv. comm.). This high-velocity kinematic component results
in the highest observed line widths of $W_{80}\approx900$\,km\,s$^{-1}$ and the most negative
asymmetry values (i.e., $A\lesssim-0.2$) observed in the Teacup AGN (Figure~\ref{fig:VelMaps}). This kinematic
component has a velocity $v=-590$\,km\,s$^{-1}$ with respect to
the systemic, or $v=-740$\,km\,s$^{-1}$ with respect to the
local velocity (i.e., with respect to $v_{{\rm p}}$; see
Figure~\ref{fig:HST}) and has a line width of ${\rm
  FWHM}=720$\,km\,s$^{-1}$. These values correspond to a maximum velocity of $v_{{\rm max}}\approx1100$\,km\,s$^{-1}$, using the
commonly used definition of $v_{\rm{max}}=|v|+\frac{1}{2}$FWHM (e.g.,
\citealt{Rupke05a,Rupke05b}). Both HR-A and HR-B
have steep radio spectra, with $\alpha_{7}^{5}=-1.13\pm0.06$ and $-1.3\pm0.2$,
respectively (see Table~\ref{Tab:RadioFluxes}). In the following sub section we explore the possible
origins of these two structures and their connection to the ionized
gas kinematics.

\subsection{What powers the core radio emission and outflow?}
\label{Sec:Core}

In this subsection we discuss which processes may be producing the radio emission in the core of the
Teacup AGN (see Figure~\ref{fig:RadioHighRes} and Figure~\ref{fig:RadioUHRes}) and what is responsible for driving the high-velocity
ionized outflow associated with the ``HR-B'' radio structure
(see Figure~\ref{fig:VelMaps} and Figure~\ref{fig:HST}). 

\subsubsection{Star formation}
Star formation is a possible candidate to produce both the radio
emission in the core and also drive outflows via stellar winds and
supernovae. The radio luminosity in the core of the Teacup AGN is
$L_{{\rm 1.4\,GHz}}=2\times10^{23}$\,W\,Hz$^{-1}$, which is
close to the proposed luminosity boundary between star-formation
dominated and AGN-dominated radio emission in low redshift quasars (i.e., $L_{{\rm 1.4\,GHz}}\approx3\times10^{23}$\,W\,Hz$^{-1}$ for $0.2<z<0.3$;
\citealt{Kimball11b}; \citealt{Condon13}). The Teacup AGN has a FIR
luminosity due to star formation of $L_{{\rm
    IR,SF}}=3\times10^{44}$\,erg\,s$^{-1}$ (\citealt{Harrison14b}). Assuming that the
majority of this FIR emission is found in the central few
kiloparsecs, the radio flux in the core region
is a factor of $\approx6$ above the radio--FIR
correlation of star-forming galaxies (e.g., \citealt{Helou85};
\citealt{Bell03}; \citealt{Ivison10b}), which would indicate that star formation
does not dominate the radio emission in the core. 

Another possibility to explore is whether or not the HR-B radio structure 
(Figure~\ref{fig:RadioHighRes}) is associated with a compact star-forming region that is responsible for driving the high-velocity gas we see co-located with
this structure (with $v=$$-740$\,km\,s$^{-1}$ and
$v_{{\rm max}}\approx1100$\,km\,s$^{-1}$;
Figure~\ref{fig:HST}). Indeed, extremely compact star
forming regions, with SFRs of $\approx$200\,M$_{\odot}$\,yr$^{-1}$
and sizes $\approx$100\,pc, have been attributed to driving 
$\approx$1000\,km\,s$^{-1}$ outflows (\citealt{Heckman11};
\citealt{DiamondStanic12a}; \citealt{Sell14}). However, maximum velocities
as high as that observed around the HR-B radio structure (i.e., $v_{{\rm max}}\approx1100$\,km\,s$^{-1}$) are
not seen in non-AGN host galaxies with SFRs comparable to that of the Teacup AGN (i.e.,
SFRs$\approx$10\,M$_{\odot}$\,yr$^{-1}$; e.g., \citealt{Genzel11};
\citealt{Arribas14}) and compact starbursts are often associated with
flatter radio spectral indices than that observed here (\citealt{Murphy13}). We further disfavor
the star formation scenario because: (1) there are not two continuum peaks observed in the {\em HST}
imaging (i.e., to correspond to both the HR-A and HR-B
radio structures); (2) HR-B is unlikely to be an obscured
star-forming region because the dust in this source is not located at this
position (see Figure~\ref{fig:HST}; also see \citealt{Keel14}); 
  (3) the high-velocity component has an [O~{\sc iii}]/H$\beta$
  emission-line flux ratio that is consistent with being dominated by AGN
  photoionization rather than a H~{\sc ii} region (see \citealt{Harrison14b}).

\subsubsection{Quasar winds}
A possible origin of the radio emission in
radio-quiet quasars is accretion disk winds (see
e.g., \citealt{Shlosman85}; \citealt{Murray95}; \citealt{Proga00};
\citealt{Crenshaw03}), which propagate through the surrounding
interstellar medium (ISM) to produce non-thermal radio emission
(\citealt{Zakamska14}; \citealt{Nims14}). Based on the energy conserving outflow model of
\cite{FaucherGiguere12b}, \cite{Nims14} explored the observational
signatures of these winds in the radio and X-rays. In their model, the wind causes a shock that propagates outwards into the ambient
medium and can reach distances of $\gtrsim$1\,kpc with velocities $\sim$1000\,km\,s$^{-1}$ (at this
point it interacts with the ISM and decelerates). They predict a steep-spectral index of $\alpha\approx-1$, which is
consistent with our observations (see Table~\ref{Tab:RadioFluxes}). It
is less clear what this model predicts for the morphology of
the radio emission; however, the propagation of the wind is predicted to be
influenced by the properties and distribution of the ISM, and it will
preferentially escape along the path of ``least resistance"
(\citealt{FaucherGiguere12b}) that could lead to a moderately
  collimated outflow. For the Teacup AGN, we find that the position of the ionized
outflow, and the associated radio emission, is directed away from the dusty regions (Figure~\ref{fig:HST}), which may be consistent with these predictions. 

In the fiducial model of \cite{Nims14}, the interaction of the winds
with the ISM is predicted to result in a radio luminosity of 
\begin{equation}
\label{Eq:Nims14}
\nu L_{\nu} \approx 10^{-5} \xi_{-2} L_{{\rm AGN}} \left(\frac{L_{{\rm
        kin}}}{0.05L_{{\rm AGN}}}\right),
\end{equation}
where $\xi_{-2}$ is the scaled efficiency
(i.e., 1\% of the shock energy goes into relativistic electrons),
$L_{{\rm AGN}}$ is the bolometric luminosity of the AGN and $L_{{\rm
    kin}}$ is the kinetic luminosity of the wind. For the Teacup
AGN, \cite{Nims14}  would predict a radio luminosity of
$\nu L_{\nu}\approx20\times10^{39}$\,erg\,s$^{-1}$ (i.e., $L_{{\rm 1.5GHz}}\approx13\times10^{23}$\,W\,Hz$^{-1}$), using their fiducial
values (i.e., $\xi_{-2}=1$; $L_{{\rm kin}}=0.05L_{{\rm AGN}}$). This
is a factor of $\approx$8 above the core radio luminosity, that we observe, of $\nu L_{\nu}{\rm
  (1.5\,GHz)}=2.6\times10^{39}$\,erg\,s$^{-1}$ (i.e., $L_{{\rm 1.5GHz}}=1.7\times10^{23}$\,W\,Hz$^{-1}$). Given the nature of
the simple model assumptions (e.g., the assumed properties of the ISM; see \citealt{Nims14}), we can not use this
discrepancy as strong evidence that the radio emission in the core of
the Teacup AGN, and the associated outflow, are not driven by a quasar
wind. 

\subsubsection{Compact jets and nuclear activity}
Compact radio emission (i.e., on $\lesssim$5\,kpc scales) in
radio-quiet quasars may be associated with nuclear coronal activity
(i.e., the region around the AGN accretion disk) or compact radio jets
(e.g., \citealt{Kukula98}; \citealt{Ulvestad05}; \citealt{Leipski06a};
\citealt{Laor08}). Nuclear coronal activity is predicted to produce
flat spectral indices (i.e., $\alpha\approx0$; \citealt{Laor08}),
which is not what is observed in the Teacup AGN
(Figure~\ref{fig:RadioSED}). 

In contrast, the core emission in the Teacup AGN shares many characteristics with radio-loud
(with $L_{{\rm 5\,GHz}}\gtrsim10^{26}$\,W\,Hz$^{-1}$) compact, steep spectrum
quasars, that show steep radio spectra, have jets and lobes on small
scales and exhibit broad [O~{\sc iii}] emission-line profiles (e.g., \citealt{Gelderman94}; see review in \citealt{ODea98}). In
particular, it is plausible that the HR-B radio structure is a hot spot of a radio jet, where it
interacts with the ionized gas in the local ISM and drives a high velocity
outflow (see Figure~\ref{fig:HST}). This is also analogous to previous observations
of a small number of AGN showing compact jets interacting with the
ionized gas (e.g., \citealt{Capetti99}; \citealt{Leipski06a};
\citealt{Barbosa09}; \citealt{Stockton07}).  We note that
  several radio-loud systems have also been observed to have high-velocity
  kinematic components of ionized gas, associated with radio jets,
  that are offset by several hundred km\,s$^{-1}$; furthermore, in the most
  extreme cases the {\em line widths} for these radio-loud systems can
  be even greater than observed in the Teacup AGN (i.e., FWHM$\gtrsim$1000\,km\,s$^{-1}$; e.g.,
  \citealt{Emonts05}; \citealt{Holt08}; \citealt{Nesvadba08}; \citealt{Shih13}).

\subsubsection{Summary of core region}
Overall the radio emission and outflow in the core of the Teacup AGN are consistent with being
produced by radio jets or quasar winds that are interacting with the
ISM in the host galaxy. This interaction could result in the high-velocity
ionized gas that we observe (see Figure~\ref{fig:HST} and
Figure~\ref{fig:Schematic}). The compact sizes and the
narrow angular range of the radio emitting region in the core, in particular for the
HR-B structure, may favor the jet scenario.

\subsection{What produces the $\approx$10\,kpc bubbles?}
\label{Sec:Bubbles}

The radio bubbles of the Teacup AGN (see Figure~\ref{fig:HST}) are reminiscent of several of the large-scale
($\gtrsim$1--10\,kpc) radio structures observed in local Seyferts
(i.e., low luminosity AGN), that are often associated with optical emission-line regions and disturbed
ionized gas (e.g., \citealt{Capetti96}; \citealt{Colbert96}; \citealt{Falcke98}; \citealt{Ferruit99}; \citealt{Whittle04};
\citealt{Hota06}; \citealt{Kharb06}). Furthermore, the morphology of the large-scale arc of ionized gas in the Teacup
AGN (Figure~\ref{fig:HST}) is analogous to some local starburst
galaxies and AGN that host ``super bubbles'', or arcs/filaments of optical emission lines (e.g.,
\citealt{Heckman90}; \citealt{Veilleux94,Veilleux01,Veilleux05}; \citealt{Capetti96}; \citealt{Ferruit99}; 
\citealt{Whittle04}; also see the analogous radio-loud AGN in
\citealt{vanBreugel85} and \citealt{Hatch13}). These emission-line features are often
attributed to the cooler edges of hot gas bubbles (with $\approx$10$^{6-8}$\,K) that are expanding and sweeping up
the ambient ISM material (e.g., \citealt{Heckman90};  \citealt{Capetti96};
\citealt{Greene14}). Simulations and theoretical models have shown that compact radio jets
(i.e., $\lesssim$1\,kpc), or quasar winds, can interact with an
inhomogeneous ambient ISM, transfer
energy and momentum and drive energy-conserving outflowing bubbles,
which expand and accelerate more clouds on larger scales (e.g.,
\citealt{Sutherland07}; \citealt{FaucherGiguere12b}; \citealt{Zubovas12};
\citealt{Wagner12, Wagner13}). It is therefore possible
that compact radio jets or a quasar wind, which may reside in the core
regions (see Section~\ref{Sec:Core}), are responsible for the $\approx$10--12\,kpc radio bubbles and the associated arc of
ionized gas (Figure~\ref{fig:RadioBubbles}; Figure~\ref{fig:RadioHighRes}).

An analogous object to the Teacup AGN is the local Seyfert Mrk~6,
which hosts 7.5\,kpc radio bubbles and $\approx$1\,kpc radio
jets. \cite{Kharb06} suggests that the radio jets are ultimately
responsible for the creation of the large-scale
radio bubbles. Following \cite{Kharb06}, we can estimate the particle energy (electrons and
protons; $E_{{\rm min}}$) at minimum pressure in the radio bubbles, under
the assumption of equipartition of energy between the magnetic field
and relativistic electrons and neglecting thermal pressure. We calculate these values following
equations (1) and (4) of \cite{ODea87} and obtain the relevant
constants from \cite{Pacholczyk70}; see \cite{Miley80} for a discussion on the
relative assumptions. We assume spherical regions of radius $D/2$ (see Table~\ref{Tab:RadioFluxes})
with a filling factor of $\phi=1$, upper and lower frequency cut offs of
$\nu_u=10$\,GHz and $\nu_l=10$\,MHz, a spectral
index of $\alpha=-0.9$ and use the
5.12\,GHz flux densities (Table~\ref{Tab:RadioFluxes}). The energy estimates also depend on the value $k$,
which is the ratio of relativistic proton to relativistic
electron energy. We consider a range of values of $E_{{\rm min}}$ by
adopting the range $k=1$--4000, which reflects the observed range
found for the radio-filled X-ray cavities in clusters
(\citealt{Birzan08}). Using these assumptions, we find that the minimum particle energies are
$E_{{\rm min}}\approx(1.7$--$130)\times10^{56}$\,erg and $E_{{\rm
    min}}\approx(0.70$--$54)\times10^{56}$\,erg for the eastern and
western radio bubbles, respectively. We note that these estimates
scale by $(\phi)^{3/7}$, such that if $\phi=10^{-3}$ this would result
in a factor of $\approx$20 decrease in the quoted values

If we assume that the HR-A and HR-B structures are both dominated by
radio jets with $\alpha=-1.1$ (see Table~\ref{Tab:RadioFluxes}), their combined total radio luminosity is
$2.5\times10^{40}$\,erg\,s$^{-1}$ (following Equation (1) in \citealt{ODea87})
which would result in a jet kinetic luminosity of
$2.5\times10^{42}$\,erg\,s$^{-1}$, assuming that 1\% of the
total jet energy is converted into radio luminosity (following
\citealt{Kharb06}; also see e.g., \citealt{Birzan08}). This would
require the jets to deposit their energy continuously for
$\approx$3.0--230\,Myr to produce the derived combined minimum energy of the
bubbles. If the core radio luminosity is, instead, produced by a
quasar wind with $L_{{\rm kin}}=1.3\times10^{43}$\,erg\,s$^{-1}$,
(following Equation~\ref{Eq:Nims14}; see \citealt{Nims14}) the wind would
need to deposit all of its energy over a timescale of
$\approx$0.59--46\,Myr. It is interesting to note that these are
  plausible timescales that are comparable to
the typical ages, of $\approx$10--40\,Myr, for the radio-filled X-ray cavities in
local cooling clusters that have similar sizes to the radio bubbles of the Teacup AGN
(\citealt{Birzan04}). To further assess these ideas, spectral
  ageing techniques could be used on deeper radio observations of the
  bubbles that have a better sampled frequency range than performed here (e.g., see \citealt{Harwood13} and
references there-in). Deep X-ray observations with {\em Chandra}
will also enable us to measure the temperature and distribution of the hot gas around the Teacup
AGN, establish if there is a hot gas component associated with the
radio bubbles and measure, more reliably, the total energy content
and lifetimes of the bubbles (following e.g., \citealt{Birzan04,Birzan08}; \citealt{Mingo11}; \citealt{Greene14}).
\newpage
\section{Conclusions} 
\label{Sec:Conclusions} 

We have presented VLA radio imaging, VIMOS/IFU optical spectroscopy and {\em HST} imaging on the Teacup
AGN; a $z=0.085$ type~2 radio-quiet quasar. Our main conclusions are
as follows.
\begin{itemize}
\item The Teacup AGN hosts bi-polar radio bubbles extending $\approx$10--12\,kpc each side
  of the nucleus (see Section~\ref{Sec:RadioData}). The edge of the brighter, eastern bubble is spatially
  coincident with an arc of luminous ionized gas (see
  Section~\ref{Sec:IFUdata} and Section~\ref{Sec:HSTdata}).
\item High-resolution radio imaging of the central few kiloparsecs reveals
  that the core is dominated by two unresolved
  radio structures (at a resolution of HPBW$\approx$0.6\,kpc). These
  are a bright central structure and a fainter structure
  $\approx$0.8\,kpc to the northeast (see Section~\ref{Sec:RadioData}). This fainter structure is
co-spatial with a high-velocity ionized gas component
($v=-740$\,km\,s$^{-1}$; $v_{{\rm max}}\approx1100$\,km\,s$^{-1}$). This high-velocity kinematic component and
corresponding radio structure are located at the base of the $\approx$12\,kpc
eastern bubble (see Section~\ref{Sec:Description}). 
\item We favor an interpretation where small-scale radio jets, or
  possibly quasar winds, are directly accelerating the gas on
  $\approx$1\,kpc scales (see Section~\ref{Sec:Core}). The same
  jets or winds, are likely to be driving the large-scale radio bubbles
  that are interacting with the ISM on $\approx$10--12\,kpc scales
  (see Section~\ref{Sec:Bubbles}). 
\end{itemize}

The Teacup AGN exhibits many of the predicted features of a source that is
the transition between an active phase of star formation and AGN activity
to a quiescent elliptical galaxy via AGN feedback (following e.g.,
\citealt{Hopkins06}), i.e., a post-merger galaxy that hosts an AGN
that is injecting energy into the gas in the host galaxy. The {\em HST}
imaging reveals that the Teacup AGN resides in a bulge-dominated galaxy, with shell-like features, indicative of previous merger activity
(Figure~\ref{fig:HST}; also see \citealt{Keel14}). Additionally, the dust lanes
in this galaxy also imply it was associated with a gas-rich merger
(e.g., \citealt{Shabala12}). This source is currently hosting quasar activity
(\citealt{Harrison14b}) that may be in decline (\citealt{Gagne14}) and our new observations
have revealed high-velocity ionized gas and radio bubbles that imply
that considerable energy injection into the ISM is occurring on $\approx$1--12\,kpc
scales, indicative of AGN feedback.

Our observations of the Teacup AGN shed some light on the observed
correlation between radio luminosity and the ionized gas kinematics (i.e., outflow velocities)
seen in radio-quiet AGN (\citealt{Mullaney13};
\citealt{Zakamska14}). The radio emission in this radio-quiet AGN is tracing both a
$\approx$1\,kpc high-velocity outflow and $\approx$10--12\,kpc radio
bubbles that are likely to both be AGN driven. In a future paper
we will present high-resolution radio imaging of a larger sample of radio-quiet
AGN that exhibit kiloparsec-scale ionized outflows to investigate the properties of their radio
emission and to determine the role, or lack there-of, of radio jets in
driving outflows and contributing to AGN feedback in these systems.

 
\noindent \smallskip\newline We thank the referee, William C. Keel,
  for his constructive comments. We acknowledge the Science and Technology Facilities Council
(CMH and MTH through grant code ST/I505656/1; APT, DMA, ACE and AMS through grant code
ST/I001573/1). and the Leverhulme Trust (DMA). FEB acknowledges
support from CONICYT-Chile (Basal-CATA PFB-06/2007, FONDECYT 1141218,
``EMBIGGEN'' Anillo ACT1101), and Project IC120009 ``Millennium
Institute of Astrophysics (MAS)'' funded by the Iniciativa
Cient\'{i}fica Milenio del Ministerio de Econom\'{i}a, Fomento y
Turismo. JRM acknowledges support from the University of Sheffield via its
Vice-Chancellor Fellowship scheme. The VLA is part of the National Radio Astronomy
Observatory, which is a facility of the National Science Foundation operated under cooperative agreement by
Associated Universities, Inc. Part of this work is based on
observations made with the NASA/ESA {\em Hubble Space Telescope}, obtained from the Data Archive at the Space Telescope Science Institute, which is operated by the Association of Universities for Research in Astronomy, Inc., under NASA contract NAS 5-26555.
 

\bibliographystyle{emulateapj}

 
\end{document}